# Strangulation as the primary mechanism for shutting down star formation in galaxies

Y. Peng[1,2], R. Maiolino[1,2], R. Cochrane[1,3]

[1] Cavendish Laboratory, University of Cambridge, 19 J. J. Thomson Ave., Cambridge CB3 0HE, UK
[2] Kavli Institute for Cosmology, University of Cambridge, Madingley Road, Cambridge CB3 0HA, UK
[3] Institute for Astronomy, Royal Observatory Edinburgh, Blackford Hill, Edinburgh EH9 3HJ, UK

**Local galaxies are broadly divided into two main classes, star forming (gas rich) and quiescent (passive and gas poor). The primary mechanism responsible for quenching star formation in galaxies and transforming them into quiescent and passive systems is still unclear. Sudden removal of gas through outflows[1,2,3,4,5,6] or stripping[7,8,9] is one of the often proposed mechanisms. An alternative mechanism is so-called "strangulation" [10,11,12,13,14], in which the supply of cold gas to the galaxy is halted. Here we report that the difference between quiescent and star forming galaxies in terms of stellar metallicity (i.e. the fraction of metals heavier than helium in stellar atmospheres) can be used to discriminate efficiently between the two mechanisms. The analysis of the stellar metallicity in local galaxies, from 26,000 SDSS spectra, clearly reveals that "strangulation" is the primary mechanism responsible for quenching star formation, with a typical timescale of 4 Gyr, at least for local galaxies with a stellar mass of $M_{star}<10^{11}$ $M_\odot$. This result is further supported independently by the stellar age difference between quiescent and star forming galaxies, which indicates that quiescent galaxies at $M_{star} < 10^{11}$ $M_\odot$ are on average observed 4 Gyr after quenching due to strangulation.**

Figure 1 qualitatively illustrates the expected evolution of galaxies in the two quenching scenarios. A more quantitative analysis is provided later on in this paper. In the scheme of Fig.1, at $t < t_q$ the galaxy is subject to gas inflow and forms stars, therefore increasing the stellar mass and the metallicity of the gas, out of which new stars form, hence the metallicity of the newly formed stars also increases with time. The metallicity increment is modest, since inflowing gas dilutes the metal content in the interstellar medium. At $t = t_q$ quenching occurs. In the scenario of sudden gas removal (e.g. by expulsion of gas by a strong wind or strong ram pressure stripping) star formation is suddenly quenched (Fig.1a), and the galaxy evolves into a quiescent system. In this case the stellar metallicity and stellar mass ($M_{star}$) of the quiescent galaxy is the same as its star forming progenitor just before quenching. In the case of quenching by "strangulation", star formation can continue with the gas available in the galaxy, until it is completely used up (Fig.1b). During this phase the gas metallicity increases more steeply than in the previous case because of the lack of dilution from inflowing gas. The stellar mass also increases slightly. The general product of strangulation is a quiescent galaxy with a stellar metallicity significantly higher than its star forming progenitor, and slightly higher stellar mass.

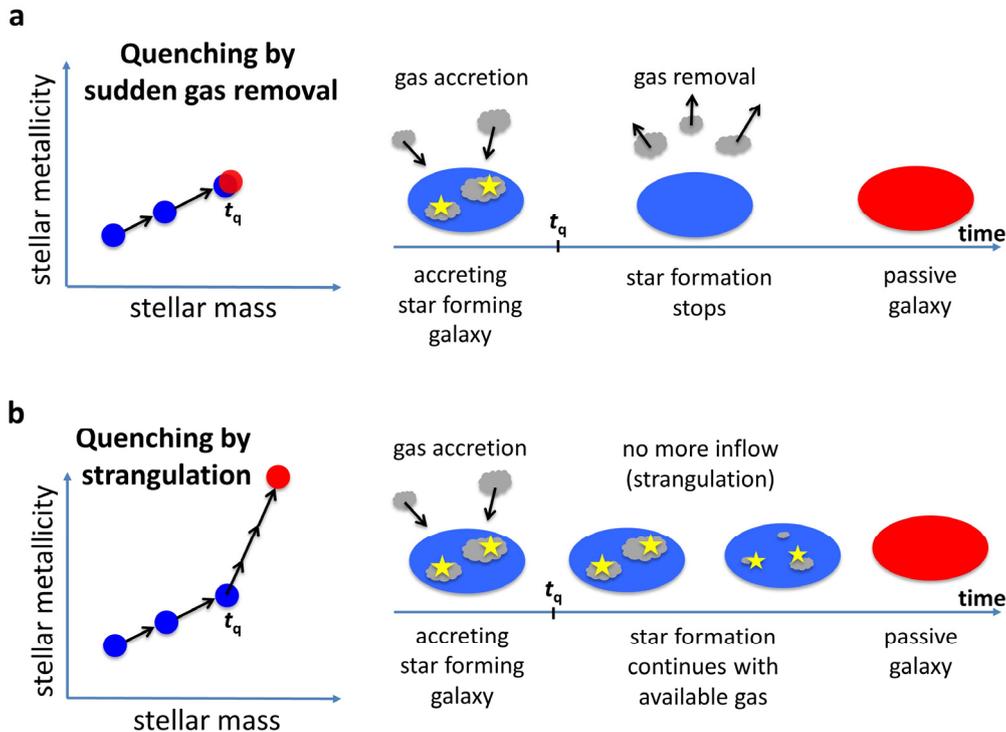

**Figure 1. Illustration of two different quenching scenarios and their effect on stellar metallicities. a,** Rapid and complete removal of the gas reservoir of the galaxy (e.g. from strong outflows or ram pressure stripping) results in a passive galaxy with the same mass and same stellar metallicity as its star forming progenitor. **b,** In the strangulation scenario, the galaxy can continue to form stars from the available enriched interstellar medium and, as a consequence, increase its stellar mass and stellar metallicity.

Observationally, obviously we cannot follow the evolution of the stellar metallicity of individual galaxies. However, we can statistically investigate the metallicity difference between star forming and quiescent galaxies. This statistical approach has already been successfully exploited, for instance, to investigate the dependence of the quenching mechanism on mass and environment[15]. We have used the Sloan Digital Sky Survey (SDSS) spectra of local ($z < \sim 0.1$) galaxies to extract a subsample of 3,905 star forming and 22,618 quiescent galaxies whose spectra have S/N>20 per spectral pixel, which ensures reliable determination of the stellar metallicities (details on the sample selection and determination of the stellar metallicities are given in the Methods).

Figure 2a shows the average stellar metallicity of star forming (blue line) and quiescent (red line) galaxies as a function of stellar mass, obtained by using a sliding average of 0.2 dex in $M_{star}$ (error bars give the 1σ uncertainty of the mean stellar metallicity). At a given stellar mass, the stellar metallicity of quiescent galaxies is significantly higher than for star forming galaxies, at least for $M_{star} < 10^{11} M_\odot$. This is not what is expected in the case of sudden gas removal, but it is qualitatively consistent with the strangulation scenario. In the following we investigate more quantitatively the agreement of the data with the strangulation scenario.

For a system forming stars without any inflow, the temporal evolution of the gaseous and stellar metallicity, as well as of the stellar mass, can be trivially solved analytically, as discussed in the Methods. The key parameters are: (1) the total gas mass at the time of quenching, $M_{gas}(t_q)$ or, equivalently, the gas fraction $f_{gas}(t_q)$; (2) the global efficiency of star formation $\varepsilon$, defined as star formation rate (SFR) = $\varepsilon\, M_{gas}$ (where $M_{gas}$ is in this case the total gas mass, both atomic and molecular); (3) the presence of any outflowing gas that is lost (i.e. which does not falls back onto the galaxy and is not recycled), which can be approximated as being proportional to the SFR and parameterized through the so-called outflow mass loading factor $\lambda$, defined as $M_{outflow} = \lambda$ SFR (note that if part of the gas falls back and is recycled[16,17], then $\lambda$ would account only for the fraction of gas that is lost, i.e. it is an "effective" outflow loading factor).

Since we are dealing with differential quantities (in particular differential stellar metallicities $\Delta Z_{star}$) the metallicity of the gas and stars at the beginning of the quenching ($t = t_q$) is unimportant. Yet, the value of $M_{star}$ at $t = t_q$ is relevant since it defines the gas fraction, hence the gas mass available for further star formation. Indeed, it is observationally well known that the gas fraction in star forming galaxies decreases with $M_{star}$[18,19,20,21] (see Methods for the detailed functional form). The global star formation efficiency $\varepsilon$ is also known from observations (see Methods).

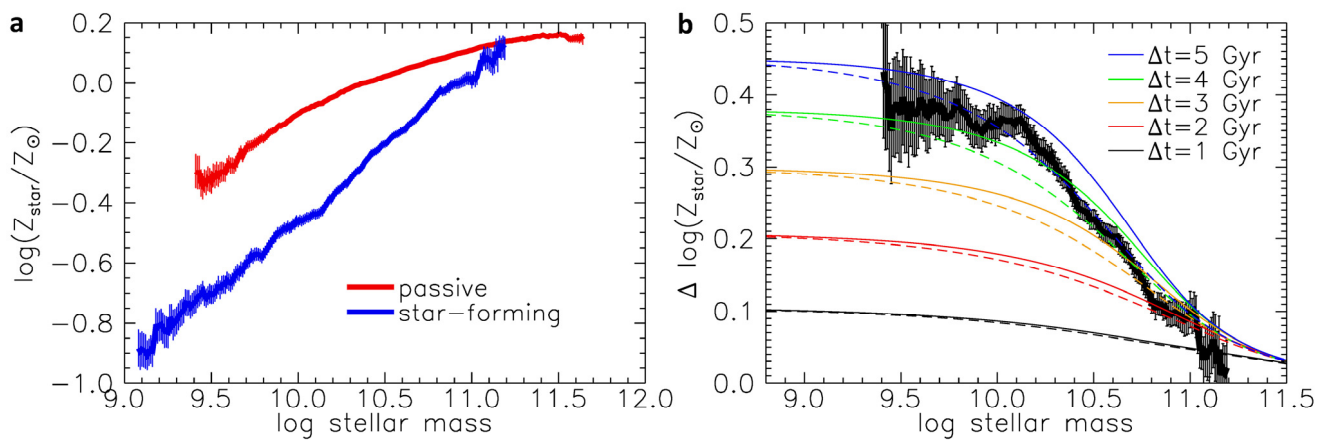

**Figure 2. Stellar metallicities for star-forming and quiescent galaxies. a,** Average stellar metallicity as a function of stellar mass for all star-forming galaxies (thick blue line with error bars) and all quiescent galaxies (thick red line with error bars) for galaxies at <$z$> ~ 0.05. Error bars correspond to the 1σ error on the mean value. **b,** Average metallicity difference between all star-forming and all quiescent galaxies (thick black line with error bars). Error bars on the black line indicate the 1σ uncertainty in the metallicity difference. The metallicity difference decreases with increasing stellar mass. It reaches the maximum value around 0.4 dex for galaxies at $M_{star}$ ~$10^{9.5} M_\odot$ and becomes negligible at $M_{star} \geq 10^{11} M_\odot$. The colored lines show the metallicity difference predicted by a simple close-box model at different times $\Delta t$ after strangulation. Solid lines are for the final mass (at time $t = t_q + \Delta t$), while dashed lines are for the mass at strangulation ($t = t_q$). The observed mass-dependent metallicity difference between quiescent and star-forming galaxies (thick black line) can be very well reproduced by a close-box model with a constant $\Delta t \sim 4$ Gyr, largely independent of stellar mass.

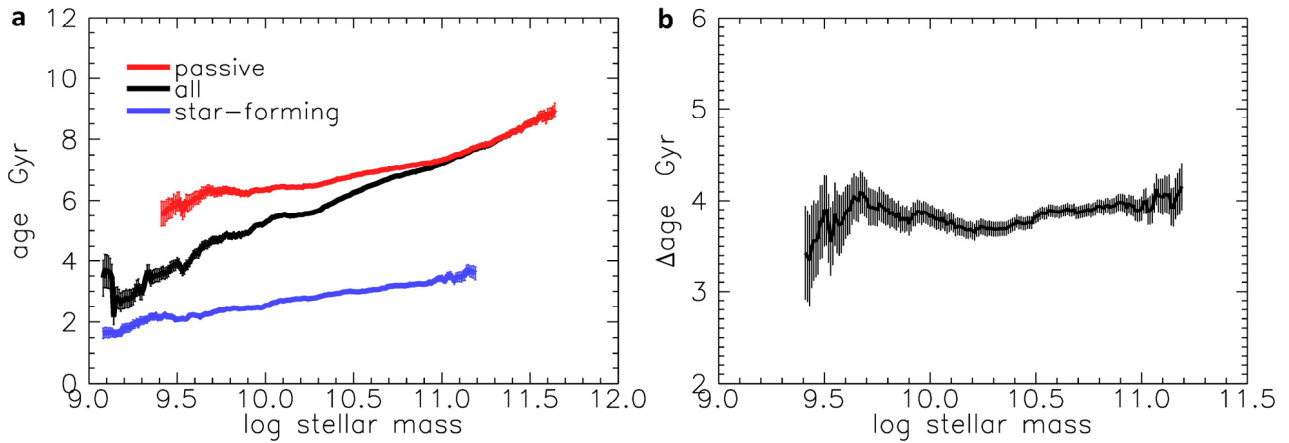

**Figure 3**. **Stellar ages for star-forming and quiescent galaxies. a,** Luminosity weighted stellar age as a function of stellar mass for star-forming (blue line), quiescent (red line) and all galaxies (black line) at $<z> \sim 0.05$. Error bars are the 1σ error on the mean value. The average age for all galaxies (black line) strongly depends on the stellar mass, largely due to the fact that the red fraction (i.e. the mixture of quiescent and star-forming galaxies) strongly depends on stellar mass. However, the dependence on stellar mass becomes much weaker once the whole sample is split into star-forming and quiescent galaxies. **b,** Average age difference between quiescent and star-forming galaxies as a function of stellar mass. Error bars on the black line indicate the 1σ uncertainty in the age difference. Remarkably, the age difference for all galaxies is largely independent of mass, with a mean value of around 4 Gyr, which is consistent with the mass-independent time Δ$t$ from strangulation required to explain the difference of stellar metallicities.

Figure 2b shows the stellar metallicity difference, as a function of stellar mass (solid thin lines in colors), as expected from the strangulation scenario, at five different times after the quenching/strangulation time ($\Delta t = t - t_q$). In this plot we are assuming that the effective outflow loading factor is $\lambda$=0 (i.e. any outflowing gas falls back and is recycled), implying that the system behaves as a closed box after strangulation (no inflow and no effective outflow). The case of significant outflow after strangulation will be discussed later on. The decline of the curves at high $M_{star}$ is primarily a consequence of the gas fraction of star forming galaxies decreasing as a function of $M_{star}$: massive galaxies have small gas content, hence once strangled the available gas can produce few stars relative to those already present, and the average stellar metallicity is little affected. The thick black line with error bars shows the observed stellar metallicity difference between quiescent and star forming galaxies (i.e. the difference between red and blue data in Fig.2a). The observed difference is consistent, within uncertainties, with the strangulation scenario in which quiescent galaxies at $M_{star} < 10^{11} M_\odot$ are, on average, observed 4 Gyr after quenching due to strangulation.

This result can be tested independently by looking at the stellar age difference between quiescent and star forming galaxies, which is measured from the galaxy spectra, independently of the stellar metallicities (see Methods). Fig.3a shows the average age of all galaxies (black line), as a function of stellar mass. The average age steeply increases with $M_{star}$, but this is mostly due to the fact that the fraction of quiescent-old galaxies, relative to star forming-young galaxies, increases with stellar mass. If the population of galaxies is split into quiescent and star forming, then the average ages of the two samples show a much flatter dependence on $M_{star}$, as illustrated by the red (quiescent) and blue (star forming) lines in Fig.3a. Most importantly, as illustrated in Fig.3b, the age difference between the two populations is reasonably constant and equal to about 4 Gyr. The latter is the same age difference required to explain the stellar metallicity difference in the strangulation scenario, therefore further and independently confirming this scenario. This relatively long timescale of ~4 Gyr is also supported by independent observations and simulations[22,23,24,25].

We also investigate the case of a significant "effective" outflow, after strangulation, by setting $\lambda$=1, which is a typical loading factor observed in star forming galaxies[26,27,28]. Fig.4 shows the resulting $\Delta Z_{star}$ curves, which are completely inconsistent with the observed data. This further confirms that gas removal by outflows plays a minor role in quenching galaxies. This includes any external environmental effect such as gas removal in satellite galaxies when falling into a more massive halo, or feedback process such as AGN driven outflow.

Overall our results strongly support the scenario in which local quiescent galaxies with $M_{star} < 10^{11} M_\odot$ (i.e. the vast majority of galaxies) are primarily quenched as a consequence of "strangulation". However, this analysis does not clarify what the strangulation mechanism is (e.g. hot halo environmental strangulation or via various preventive feedback mechanisms, such as circumgalactic gas heating[29]). Additional, more difficult analysis has to be performed to investigate the strangulation mechanism (e.g. by studying the central/satellite and environmental dependence as briefly discussed in Section 3 in Methods). We also note that our results do not imply any claim on the morphological changes of the galaxy population, as it is not completely clear whether the morphological transformation is associated with star-formation quenching[30].

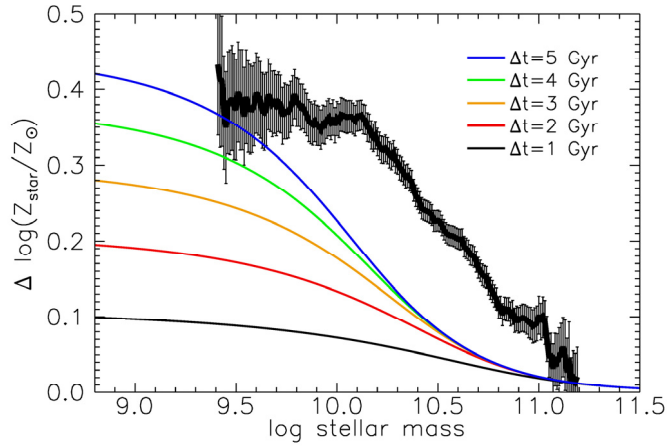

**Figure 4. The effect of outflows on stellar metallicity evolution.** As Fig.2b, but for the case of an effective mass-loading factor $\lambda=1$ after strangulation. In this case the observed stellar metallicity difference (black thick line) is clearly not reproduced by the model, further suggesting that outflows do not play a major role. Error bars on the black line indicate the 1σ uncertainty in the metallicity difference.

The data presented in this paper cannot shed light on the quenching mechanism at $M_{star} \geq 10^{11} M_\odot$. At $M_{star} \sim 10^{11} M_\odot$ the stellar metallicity of quenched galaxies is similar to the stellar metallicity of star forming galaxies (Fig.2), which can be interpreted equally well as quenching by sudden gas removal (e.g. by outflows, Fig.1a) or as quenching by strangulation of gas-poor massive galaxies (indeed in massive galaxies the small amount of available gas, as shown in Extended Data Fig.1a, does not allow much star formation, and hence there is little variation of the stellar metallicity, even if the galaxy is strangled. See also discussions in Section 4 in Methods). At even higher stellar masses our analysis is not feasible, since star forming galaxies with $M_{star} > 10^{11} M_\odot$ are extremely rare in the local Universe, hence preventing our statistical approach. To shed light on the quenching mechanism of massive galaxies a similar analysis has to be performed at high-$z$, where massive star forming galaxies are abundant and gas rich.

The results obtained in this paper apply *on average* to the bulk of the local galaxy population. However, for individual galaxies, other quenching mechanisms such as fast gas removal via outflows, that can also help to explain the α-element enhancement in massive ellipticals, and environmental effects (e.g. ram-pressure stripping, tidal stripping, harassment and mergers), may work together with (or actually cause) strangulation to shape the detailed quenching process (see Section 4 in Methods). These additional quenching mechanisms may modify the amount of stellar metallicity enhancement, the quenching timescale, and/or the gas content, which can all contribute to the scatter in the stellar metallicity and age differences between star-forming and quiescent galaxies.

**References:**

1. Di Matteo, T, Springel, V. & Hernquist, L. Energy input from quasars regulates the growth and activity of black holes and their host galaxies. *Nature*. **433**, 604-607 (2005)
2. Hopkins, P. F., Hernquist, L., Cox, T. J., Di Matteo, T., Robertson, B. & Springel, V. A Unified, Merger-driven Model of the Origin of Starbursts, Quasars, the Cosmic X-Ray Background, Supermassive Black Holes, and Galaxy Spheroids. *Astrophys. J. Suppl. Ser.* **163**, 1-49 (2006)
3. Maiolino, R. et al. Evidence of strong quasar feedback in the early Universe. *Mon. Not. Royal Astron. Soc*. **425**, L66-L70 (2012)
4. Diamond-Stanic, A. et al. High-velocity Outflows without AGN Feedback: Eddington-limited Star Formation in Compact Massive Galaxies. *Astrophys. J.* **755**, L26 (2012)
5. Förster Schreiber, N. M. et al. The Sins/zC-Sinf Survey of $z \sim 2$ Galaxy Kinematics: Evidence for Powerful Active Galactic Nucleus-Driven Nuclear Outflows in Massive Star-Forming Galaxies. *Astrophys. J.* **787**, 38 (2014)
6. Cicone, C. et al. Massive molecular outflows and evidence for AGN feedback from CO observations. *Astron. Astrophys.* **562**, A21 (2014)
7. Gunn, J.E. & Gott, J.R. On the Infall of Matter Into Clusters of Galaxies and Some Effects on Their Evolution. *Astrophys. J.* **176**, 1-19 (1972)
8. Abadi, M. G., Moore, B. & Bower, R. G. Ram pressure stripping of spiral galaxies in clusters. *Mon. Not. Royal Astron. Soc*. **308**, 947-954 (1999)
9. Quilis, V., Moore, B. & Bower, R. Gone with the Wind: The Origin of S0 Galaxies in Clusters. *Science*. **288**, 1617-1620 (2000)
10. Larson, R. B., Tinsley, B. M. & Caldwell, C. N. The evolution of disk galaxies and the origin of S0 galaxies. *Astrophys. J.* **237**, 692-707 (1980)
11. Balogh, M. L. & Morris, S. L. Hα photometry of Abell 2390. *Mon. Not. Royal Astron. Soc.* **318**, 703-714 (2000)

12. Balogh, M. L., Navarro, J. F. & Morris, S. L. The Origin of Star Formation Gradients in Rich Galaxy Clusters. *Astrophys. J.* **540**, 113-121 (2000)
13. Kereš, D., Katz N., Weinberg, D. H. & Davé, R. How do galaxies get their gas? *Mon. Not. Royal Astron. Soc.* **363**, 2-28 (2005)
14. Dekel, A. & Birnboim, Y. Galaxy bimodality due to cold flows and shock heating. *Mon. Not. Royal Astron. Soc.* **368**, 2-20 (2006)
15. Peng, Y. et al. Mass and Environment as Drivers of Galaxy Evolution in SDSS and zCOSMOS and the Origin of the Schechter Function. *Astrophys. J.* **721**, 193-221 (2010)
16. De Lucia, G., Kauffmann, G. & White, S. D. M. Chemical enrichment of the intracluster and intergalactic medium in a hierarchical galaxy formation model. *Mon. Not. Royal Astron. Soc.* **349**, 1101-1116 (2004)
17. Oppenheimer, B. D. et al. Feedback and recycled wind accretion: assembling the $z = 0$ galaxy mass function. *Mon. Not. Royal Astron. Soc.* **406**, 2325-2338 (2010)
18. Baldry, I. K., Glazebrook, K. & Driver, S. P. On the galaxy stellar mass function, the mass-metallicity relation and the implied baryonic mass function. *Mon. Not. Royal Astron. Soc.* **388**, 945-959 (2008)
19. Peeples, M. S. & Shankar, F. Constraints on star formation driven galaxy winds from the mass-metallicity relation at $z=0$. *Mon. Not. Royal Astron. Soc.* **417**, 2962-2981 (2011)
20. Boselli, A., Cortese, L., Boquien, M., Boissier, S., Catinella, B., Lagos, C. & Saintonge, A. Cold gas properties of the Herschel Reference Survey. II. Molecular and total gas scaling relations. *Astron. Astrophys*, **564**, A66 (2014)
21. Santini, P. et al. The evolution of the dust and gas content in galaxies. *Astron. Astrophys.* **562**, A30 (2014)
22. Wetzel, A. R., Tinker, J. L., Conroy, C. & van den Bosch, F. C. Galaxy evolution in groups and clusters: satellite star formation histories and quenching time-scales in a hierarchical Universe. *Mon. Not. Royal Astron. Soc.* **432**, 336-358 (2013)
23. Hirschmann, M. et al. The influence of the environmental history on quenching star formation in a Λ cold dark matter universe. *Mon. Not. Royal Astron. Soc.* **444**, 2938-2959 (2014)
24. Schawinski, K. et al. The green valley is a red herring: Galaxy Zoo reveals two evolutionary pathways towards quenching of star formation in early- and late-type galaxies. *Mon. Not. Royal Astron. Soc.* **440**, 889-907 (2014)
25. Woo, J., Dekel, A., Faber, S. M. & Koo, D. C. Two conditions for galaxy quenching: compact centres and massive haloes. *Mon. Not. Royal Astron. Soc.* **448**, 237-251 (2015)
26. Davé, R., Finlator, K. & Oppenheimer, B. D. Galaxy evolution in cosmological simulations with outflows - II. Metallicities and gas fractions. *Mon. Not. Royal Astron. Soc.* **416**, 1354-1376 (2011)
27. Hopkins, P. F., Quataert, E. & Murray, N. Stellar feedback in galaxies and the origin of galaxy-scale winds. *Mon. Not. Royal Astron. Soc.* **421**, 3522-3537 (2012)
28. Cicone, C. et al. Massive molecular outflows and evidence for AGN feedback from CO observations. *Astron. Astrophys*, **562**, A21 (2014)
29. Davé, R., Finlator, K. & Oppenheimer, B. D. An analytic model for the evolution of the stellar, gas and metal content of galaxies. *Mon. Not. Royal Astron. Soc.* **421**, 98-107 (2012)
30. Carollo, C. M. et al. ZENS IV. Similar Morphological Changes associated with Mass- and Environment-Quenching, and the Relative importance of Bulge Growth versus the Fading of Disks. *Astrophys. J.* submitted (arXiv: 1402.1172) (2014)

**Acknowledgments:** We are very grateful to Anna Gallazzi and her collaborators for making publicly available of their SDSS DR4 version of the stellar ages and metallicities catalogues. We thank Simon Lilly, Alvio Renzini, Hans-Walter Rix and Martin Haehnelt for helpful discussions. We acknowledge NASA's IDL Astronomy Users Library, the IDL code base maintained by D. Schlegel, and the *k*correct software package of M. Blanton.

**Authors Contributions** Y.P. and R.M. co-developed the idea, both contributed to the interpretation and manuscript writing. Y.P. and R.C. contributed to the measurements and analysis.

**Author Information** Reprints and permissions information is available at www.nature.com/reprints. The authors declare no competing financial interests. Readers are welcome to comment on the online version of the paper. Correspondence and requests for materials should be addressed to Y.P. (yp244@mrao.cam.ac.uk)

**Methods**

**1. Sample and observational data**

The parent sample of galaxies analyzed in this paper is the same SDSS DR7 sample used in ref.15 & 31 for similar statistical investigations of the quenching process. Briefly, it is a magnitude-selected sample of galaxies in the redshift range of $0.02 < z < 0.085$ that have clean photometry and Petrosian SDSS $r$-band magnitudes in the range of $10.0 < r < 18.0$ after correcting for Galactic extinction. The parent photometric sample contains 1,579,314 objects after removing duplicates, of which 238,474 have reliable spectroscopic redshift measurements. As a consequence of the relatively broad redshift range of $0.02 < z < 0.085$, the projected physical aperture of the SDSS spectroscopic fiber changes significantly across the sample. To test whether there are any significant aperture effects, we further split the whole redshift range into two narrower redshift ranges, at $0.02 < z < 0.05$ and $0.05 < z < 0.085$. The results are shown in the Extended Data Fig.2. It is clear that the results do not change significantly as a function of redshift.

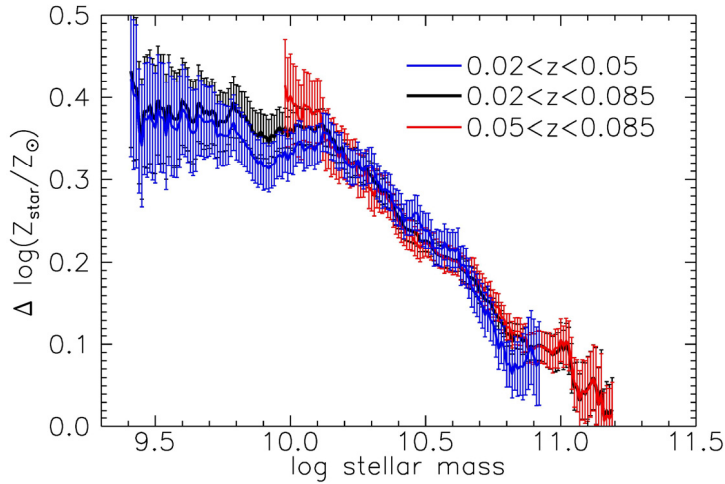

**Extended Data Fig.2. Stellar metallicity difference in different redshift bins.** To investigate any aperture effects, the sample over the whole redshift range of $0.02 < z < 0.085$ is further divided into two narrower redshift ranges of $0.02 < z < 0.05$ and $0.05 < z < 0.085$. It is clear that the results do not change significantly as a function of redshift, i.e. as a function of projected aperture. The error bars on each line indicate the $1\sigma$ uncertainty in the metallicity difference.

Each galaxy is weighted by $1/TSR \times 1/V_{max}$, where $TSR$ is a spatial target sampling rate, determined using the fraction of objects that have spectra in the parent photometric sample within the minimum SDSS fiber spacing of 55 arcsec of a given object. The $V_{max}$ values are derived from the $k$-correction program v4_1_4[32]. The use of $V_{max}$ weighting allows us to safely include representatives of the galaxy population down to a stellar mass of about $10^9\ M_\odot$.

The stellar masses are determined from the $k$-correction code with the Bruzual & Charlot (2003, ref.33) population synthesis models and a Chabrier initial mass function (IMF). The galaxy population is then divided into star-forming and passive galaxies, based on their spectroscopic emission line classifications and rest-frame (U−B) colors. Star-forming galaxies are defined to have the classification flag (i.e. the "iclass" keyword), in Brinchmann et al. (2004, ref.34) SFR catalogue, set to 1. This also excludes those objects hosting an active galactic nucleus (AGN), for which the derived metallicities are probably not reliable. Passive galaxies are defined to have their (U−B) colors redder than a threshold color that is given by Equation 2 in ref.15 and without Hα emission (i.e. undetected with a signal-to-noise ratio (S/N) < 3).

The stellar metallicities and $r$-band weighted stellar ages were derived by Gallazzi et al. (2005, ref.35) from the spectral absorption features of SDSS DR4 spectra, which are then cross-matched with our parent galaxy sample. We refer to Gallazzi et al. (2005) for a detailed description of the method. Here, we only mention that the method consists of measuring the strength of a set of carefully selected spectral absorption features, including several well calibrated Lick indices and the 4000 Å break. Then they compute the median-likelihood estimates of the stellar metallicities and $r$-band light weighted ages by comparing the spectral absorption features to a large library of 150,000 Monte Carlo realizations spanning a full range of physically plausible star formation histories. The stellar metallicities are derived for both passive and star-forming galaxies, for which the contamination of stellar absorption features by nebular emission has been carefully removed. To ensure a reliable metallicity measurement, we restrict our galaxy sample to galaxies that have a median S/N per pixel of at least 20 over the whole spectrum. This ensures the average uncertainty of the stellar metallicity and light-weighted age to be less than ±0.15 dex. The requirement on the S/N could, in principle, be lowered in order to increase the statistics, but this would result in higher uncertainties in the metallicity and age measurements. Since the statistical errors on the mean metallicity and mean age are already reasonably small with the above S/N selection, the analysis benefits more from solid measurements than from improved statistics.

With the selection criteria given above, the final sample consists of 22,618 passive galaxies and 3,905 star-forming galaxies. All of these galaxies have reliable stellar mass, stellar metallicities and age measurements.

## 2. Metallicity evolution during quenching via strangulation

The quenching process via strangulation can be quantitatively described by using the analytical framework discussed in ref.36&37. These simple analytical models take into account the key physical processes of inflow, star formation, outflow and metal production.

We assume the star formation law holds in the same way before and during the quenching process, i.e. the instantaneous average SFR of the galaxy is always related to the gas mass present within the galaxy as

$$\text{SFR} = \varepsilon M_{\text{gas}} \quad (1)$$

where $M_{\text{gas}}$ is the total gas mass (both atomic and molecular). In fact, Equation (1) can be regarded as the definition of effective, global star-formation efficiency $\varepsilon$ or, more properly, inverse of the total gas depletion timescale $\tau_{\text{dep}} = M_{\text{gas}}/\text{SFR} = 1/\varepsilon$.

Note that, from Equation (1), the specific SFR (sSFR) can be expressed as: $\text{sSFR} = \text{SFR}/M_{\text{star}} = \varepsilon M_{\text{gas}}/M_{\text{star}}$.

The mass-loss rate of the galaxy $\Psi$, i.e. the outflow rate, is very likely to be closely related to the average SFR of the galaxy. Analogous to Equation (1), we link these two quantities together via $\lambda$, as

$$\Psi = \lambda \cdot \text{SFR} \quad (2)$$

where $\lambda$ is the mass-loading factor. Similar to $\varepsilon$, Equation (2) can be regarded as the definition of $\lambda$. It should be noted that strangulating the gas inflow does not necessarily turn the gas regulator model to a simple close-box model, since the galaxy can continue to have an outflow.

The general evolution of the gas metallicity $Z_{\text{gas}}$, without assuming any equilibrium condition, is given by Equation 32 in ref.37, i.e.

$$\frac{dZ_{\text{gas}}}{dt} = y\varepsilon - (Z_{\text{gas}} - Z_0)\frac{\Phi}{M_{\text{gas}}} \quad (3)$$

where $y$ is the average yield per stellar generation and is assumed to be a constant, $\Phi$ is the inflow rate, $Z_0$ is the metallicity of the infalling gas and $M_{\text{gas}}$ is the gas mass. When quenching via strangulation starts, $\Phi$ is set to zero and Equation (3) can be easily solved, as

$$Z_{\text{gas}}(t) = Z_{\text{gas}}(t_q) + y\varepsilon t \quad (4)$$

where $Z_{\text{gas}}(t_q)$ is the gas metallicity at the time when quenching begins. It is clear from Equation (4) that $\Delta Z_{\text{gas}} \sim y\varepsilon t$, i.e. for a constant yield and star-formation efficiency, the gas metallicity increase is simply proportional to time. It is also interesting to notice that *when the inflow is truncated, the gas metallicity is independent of the outflow*.

From Equation (4), the logarithmic increase of the gas metallicity when quenching begins is given by

$$\log Z_{\text{gas}}(t) - \log Z_{\text{gas}}(t_q) = \log[1 + \frac{y\varepsilon t}{Z_{\text{gas}}(t_q)}] \quad (5)$$

Before the start of the quenching, according to Equation 35 in ref.37, $Z_{\text{gas}}$ is proportional to the yield $y$ if $Z_0 \sim 0$. By inserting it into the right hand side of Equation (5), $y$ will cancel out from both the numerator and denominator. Therefore, *when the inflow is truncated, the amount of logarithmic increase of the gas metallicity is independent of the yield*. Hence, any uncertainty on the yield is completely unimportant for our analysis.

The general evolution of the stellar metallicity $Z_{\text{star}}$, without assuming any equilibrium condition, is given by Equation 40 in ref.37 as

$$\frac{dZ_{\text{star}}}{dt} = \text{sSFR} \cdot (1-R)(Z_{\text{gas}} - Z_{\text{star}})$$
$$= \frac{\varepsilon M_{\text{gas}}}{M_{\text{star}}}(1-R)(Z_{\text{gas}} - Z_{\text{star}}) \quad (6)$$

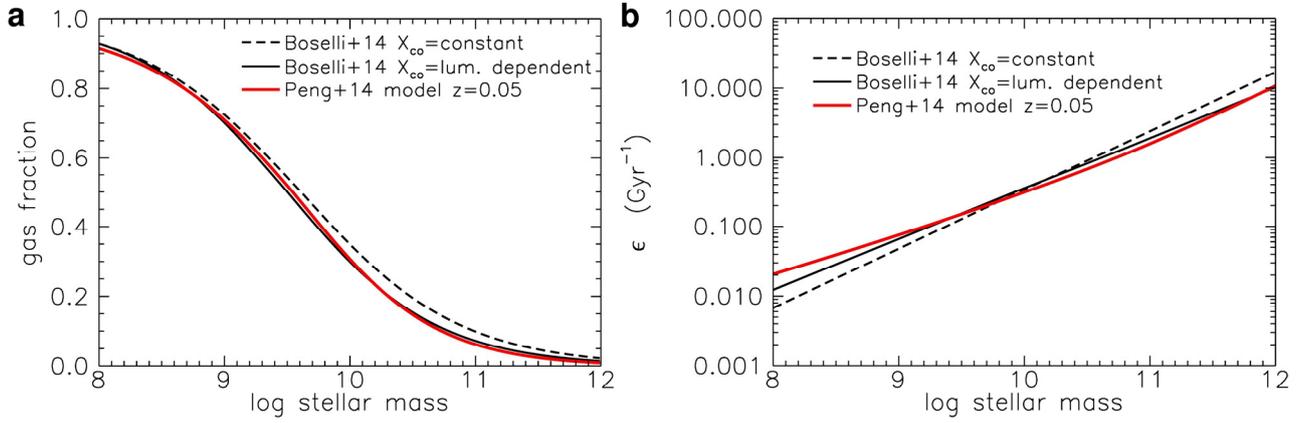

**Extended Data Fig.1**. **Gas fraction and star-formation efficiency for star-forming galaxies. a,** The observed total gas fraction for local galaxies determined using a constant CO-to-$H_2$ conversion factor $X_{CO}$ (black dashed line) and an H-band luminosity-dependent conversion factor (black solid line) in ref. 20. The predicted total gas fraction (molecular and atomic) for star-forming galaxies as a function of stellar mass from the Peng et al. (2014, ref.37) model at $z \sim 0.05$ (red solid line). **b,** Star-formation efficiency $\varepsilon$, defined as $\varepsilon = SFR/M_{gas}(tot)$, i.e. the reverse of the gas depletion timescale, as a function of stellar mass.

where $R$ is the fraction of the mass of the newly formed stars that is quickly returned to the interstellar medium through stellar winds and supernovae. It is clear from Equation (6) that the stellar metallicity simply evolves towards the gas metallicity on a timescale controlled by 1/sSFR.

The change of stellar mass of the galaxy per unit time is given by

$$\frac{dM_{star}}{dt} = (1-R) \cdot SFR \qquad (7)$$

where $(1-R)$SFR is the net SFR that contributes to the net stellar mass increase of the galaxy, i.e. the fraction of newly produced stars in the form of long-lived stars. The change of the gas mass of the galaxy per unit time is given by

$$\begin{aligned}\frac{dM_{gas}}{dt} &= -(1-R) \cdot SFR - \Psi \\ &= -(1-R+\lambda)\varepsilon M_{gas}\end{aligned} \qquad (8)$$

To calculate the change of the stellar metallicity during quenching, we need to know, at a given stellar mass, the gas mass (or equivalently the gas fraction), the star-formation efficiency $\varepsilon$ and the mass-loading factor $\lambda$.

For $\lambda$, we have first assumed in the paper that, during the strangulation process, the galaxy can recycle all the outflow gas by setting $\lambda=0$, i.e. we have assumed a close-box model. Then, as discussed in the text, we have also investigated the case of $\lambda=1$ during strangulation.

Both gas fraction and star-formation efficiency $\varepsilon$ (or equivalently the gas depletion timescale $\tau_{dep}$) have been measured observationally[18,19,20,21]. In fact, the predicted gas fraction and $\varepsilon$ by the model in ref.37 match the latest observations in the local Universe[20] extremely well, as shown in the Extended Data Fig.1. We stress again that the gas fraction and star-formation efficiency in all our calculations are defined with the *total* gas mass (including both atomic and molecular). While in some literature such as ref.38, the average gas depletion time is found to be about constant of ~2 Gyr, which refers to the molecular gas depletion time (see further discussion in Section 5 in Methods).

The change in stellar metallicity as a function of stellar mass at different times $\Delta t$ after strangulation is shown in Fig.2b (in the case $\lambda=0$). During strangulation galaxies will continue to form stars with the available gas following the star-formation law as given by Equation (1) and their stellar mass will hence continue to grow. The colored solid lines show the stellar metallicity increase as a function of the final stellar mass, while the colored dashed lines show the stellar metallicity increase as a function of the stellar mass at the epoch when the strangulation starts. As shown in Fig.2b, the stellar metallicity increase is slightly larger if the considered stellar mass is the final stellar mass, but there are no significant differences between these two stellar masses.

At a given stellar mass, the amount of stellar metallicity increase is nearly proportional to the time interval $\Delta t$ elapsed from the beginning of strangulation. At a given $\Delta t$ the stellar metallicity increase is larger for low mass galaxies than for more massive galaxies. This is because the variation of stellar metallicity depends on the size of the gas reservoir, i.e. the gas fraction, at the epoch when the strangulation starts (the larger the gas reservoir the more metals can be produced in the strangulation phase). For massive galaxies, which have a low gas fraction (Extended Data Fig.1a), the relative amount of gas available for star formation is small, while the existing stellar population (with relatively low metallicity) is large compared to the amount of new stars (with higher metallicity) that will form during the strangulation. Therefore, although the star-formation efficiency is higher for massive galaxies (Extended Data Fig.1b), the increase in stellar metallicity is smaller for massive galaxies than for low-mass galaxies (Fig.2b).

It is evident, as discussed in the main text of the paper, that the observed metallicity difference can be very well reproduced by a simple close-box model with a constant mass-independent $\Delta t \sim 4$ Gyr across the entire observed range of stellar masses, which is consistent with the age difference between the two populations (Fig.3b).

We stress that a different stellar metallicity calibration and age calibration method may give different scales and different slopes of the stellar mass versus metallicity/age relations, but the results illustrated above are preserved regardless of the adopted calibration. This is because our results are mainly based on metallicity *differences* and age *differences* between star-forming and passive galaxies, i.e. we are dealing with differential quantities, therefore uncertainties in the metallicity/age scale are much less critical than in studies dealing with the absolute quantities.

Finally, we discuss one second-order effect that we have not taken into consideration, but which would reinforce our results even further. The star forming progenitors of passive (quenched) galaxies observed locally should be star forming galaxies at $z\sim0.5$ (i.e. 4 Gyr ago), and not the star forming population observed locally in SDSS. Unfortunately, SDSS does not have the sensitivity to deliver star forming galaxies at high redshifts in large numbers and with the same S/N as local star forming galaxies. However, star forming galaxies at $z\sim0.5$ should have a metallicity even lower than local galaxies. Therefore, if any, the metallicity difference between passive and star forming galaxies observed in Fig.2 should be even larger. However, this effect is expected to be very small, since the mass-metallicity relation evolves very little from $z=0$ to $z=0.5$[39,40].

## 3. Stellar metallicity for central and satellite galaxies

Since a distinction between central and satellite galaxies appears in many theoretical models for the evolution of galaxies and may potentially shed light on the strangulation mechanism, we further divide the whole sample into central and satellite galaxies and the results are shown in the Extended Data Fig.3. As discussed in ref.31, there are difficulties in the identification of true centrals due to over fragmentation of groups by the group finding algorithm, which would lead to some satellites being misidentified as centrals. This effect would be expected to be most severe for low-mass galaxies in high density regions. To obtain a clean sample of true centrals, we further select the centrals in the fields, where their over-densities are below the mean over-density of the local universe.

The stellar metallicity enhancement of satellites is slightly larger than that of centrals at $M_{star} < 10^{10}$ $M_\odot$. This suggests that low mass satellites are likely to suffer more from the strangulation than centrals, hence an environmental origin of strangulation at these low masses (e.g. inflow of gas being halted as satellite galaxies plunge into the hot halos). At masses $M_{star} > 10^{10}$ $M_\odot$ no difference between central and satellites is detected. This suggests that at high masses the strangulation process may operate similarly for both centrals and satellites. A more detailed analysis, in different environments, is required to unveil the origin of strangulation at high masses, which will be presented in a future work.

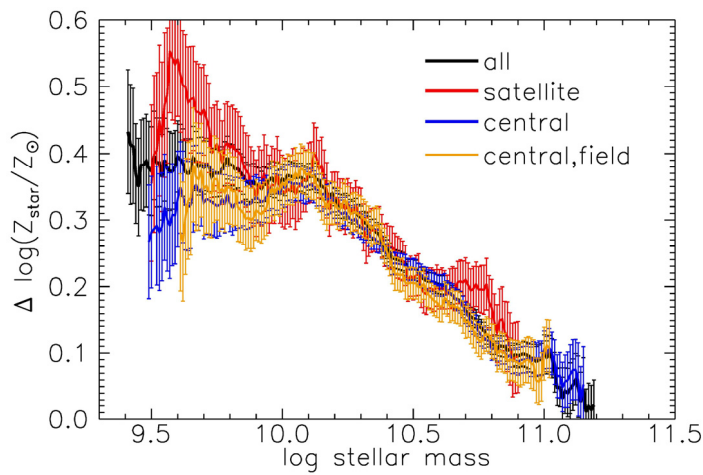

**Extended Data Fig.3**. **Stellar metallicity difference for central and satellite galaxies.** The whole sample is further divided into centrals and satellites. The orange line shows the central galaxies in the field, which represent a clean sample of true centrals as explained in the text. The stellar metallicity enhancement of satellites is slightly larger than that of centrals at $M_{star} < 10^{10}$ $M_\odot$ (suggesting that environment may play a role in the strangulation mechanism at these low masses), while no detectable difference between them is seen at higher stellar masses. The error bars indicate the 1σ uncertainty in the metallicity difference.

## 4. Fraction of galaxies quenched by rapid gas removal

We have argued in this paper that the majority of local quiescent galaxies with $M_{star} < 10^{11}$ $M_\odot$ are primarily quenched as a consequence of strangulation. This is a statistical statement, based on the average properties of the galaxy population. It is however interesting to quantify the fraction of galaxies whose data may potentially allow rapid quenching by sudden gas removal, which does not cause metallicity change, as an alternative viable process.

In each panel of the Extended Data Fig.4, we show the probability density function (PDF) of the stellar metallicity of star-forming galaxies (blue line) and that of the passive galaxies (red line) at a given stellar mass (as noted in the label). Each PDF is normalized (i.e. the area beneath each curve is unity). The overlapping region of the two PDFs is shaded (light red). The fraction of the shaded over the total area given by the blue PDF is noted as $f_{max}$ in the label.

If all star forming galaxies were eventually to be quenched, the star forming PDF (blue line) should eventually evolve into the passive PDF (red line).

We discussed that a significant difference between star forming and passive metallicity implies strangulation as primary quenching process. The opposite is not necessarily true: a similar metallicity between star forming and passive galaxies may be caused either by sudden gas removal (outflows, gas stripping, etc.) or strangulation of galaxies with modest gas content. Furthermore, we also note that metallicities of star forming galaxies similar to those of passive ones can also trace galaxies that initially had large gas content, were strangled, and we are observing them during the last stages of star formation in their strangulation phase. Therefore the shaded area, and hence $f_{max}$, give a very conservative upper limit to the fraction of galaxies for which sudden gas removal can potentially be an alternative quenching mechanism.

$f_{max}$ is 50% at low masses, confirming that the bulk of the galaxies at low masses must quench by strangulation. We recall, as discussed above, that this is a very conservative approach. $f_{max}$ progressively increases with increasing stellar mass. The latter effect is due to the fact that the observed stellar metallicity enhancement decreases with increasing stellar mass as shown in Fig.2b. As shown in that figure, the lines with different $\Delta t$ converge to zero with increasing mass. This implies that, as already discussed, the constraining power of the stellar metallicity data on the quenching mechanism decreases with increasing mass. At the massive end of $M_{star} > 10^{11}$ $M_\odot$, the stellar metallicity data cannot shed light on the quenching mechanism, as the small metallicity difference can be interpreted equally well as rapid quenching by sudden gas removal or as quenching by strangulation.

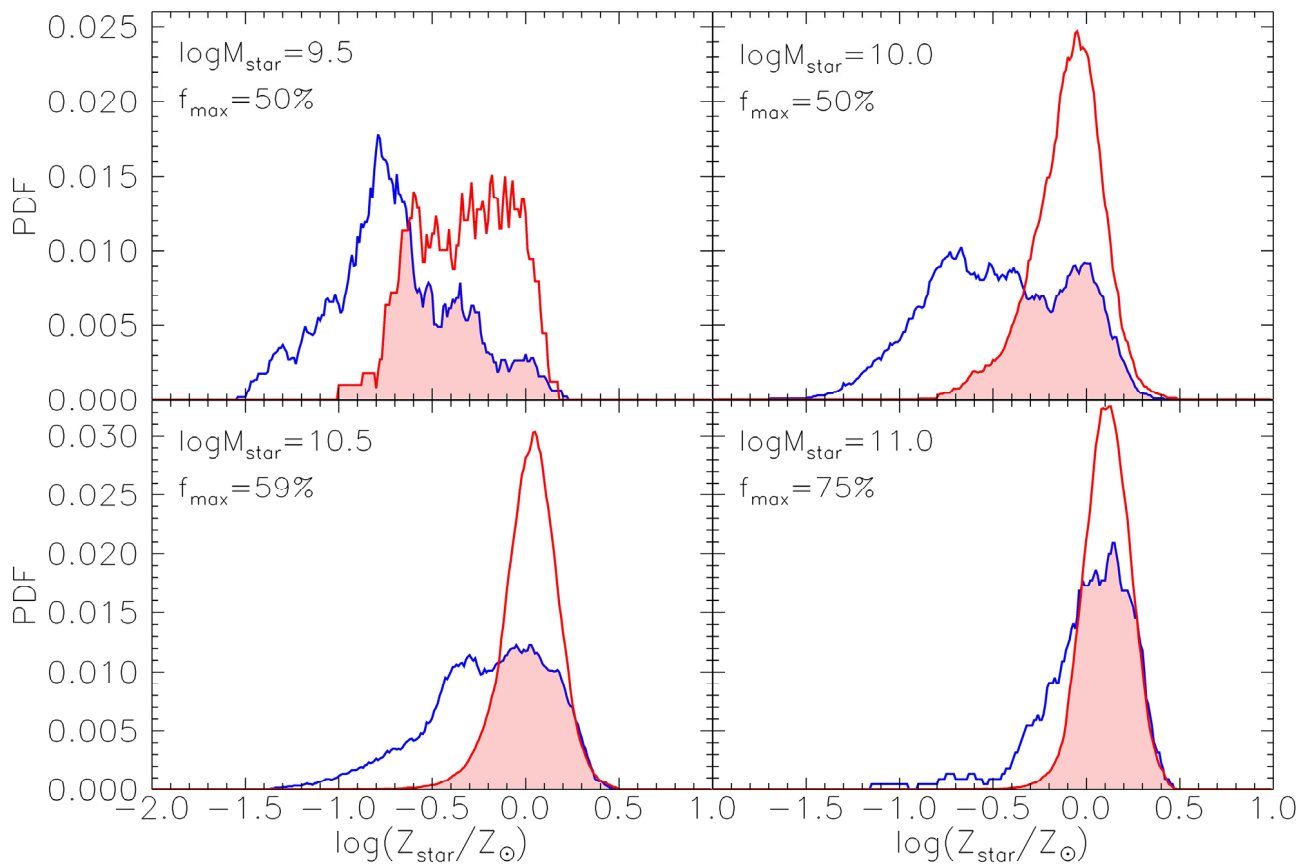

**Extended Data Fig.4**. **Probability density function of star-forming and passive galaxies.** In each panel the blue line shows the probability density function (PDF) of the stellar metallicity of star-forming galaxies at a given stellar mass (as noted in the label) and red line shows the corresponding PDF of passive galaxies. The overlapping region of the two PDFs is shaded (light red). The fraction of the shaded area over the total area given by each of PDF (noted as $f_{max}$ in the label) gives the maximum fraction of galaxies for which rapid gas removal may be an allowed alternative quenching mechanism.

## 5. Effect of constant star formation efficiency on stellar metallicity evolution

As discussed in Section 2 of the Methods, although the star formation efficiency $\varepsilon$ defined with the *total* gas mass (i.e. $\varepsilon$ = SFR/($M_{HI}+M_{H2}$)) is very unlikely to be constant, it is useful to test further the validity of the result in Fig.2b against a putative constant $\varepsilon$. We keep all other parameters unchanged in the model, except that when strangulation starts we set $\varepsilon$ to be a constant of 0.5 Gyr$^{-1}$, i.e. a constant gas depletion timescale of $\tau_{dep}$ = 2 Gyr. The results are shown in the Extended Data Fig.5. It is clear that at $M_{star}$ > ~$10^{10}$ $M_{\odot}$, the prediction from the model is very similar to Fig.2b and is still consistent with $\Delta t$ ~ 4 or 5 Gyr. This is because the average $\varepsilon$ in the Extended Data Fig.1b at $M_{star}$ > ~$10^{10}$ $M_{\odot}$ is roughly 0.5 Gyr$^{-1}$. However, below ~$10^{10}$ $M_{\odot}$, the model now predicts a much steeper metallicity increase, due to the adopted constant value of $\varepsilon$ = 0.5 Gyr$^{-1}$ that is much higher than the one shown in the Extended Data Fig.1b. This suggests that, to achieve the same observed metallicity enhancement, a higher $\varepsilon$ will need a shorter $\Delta t$. In other words, the strangulation process can be fast (e.g. less than 1 Gyr) if the $\varepsilon$ is large (for instance, at higher redshifts[21]).

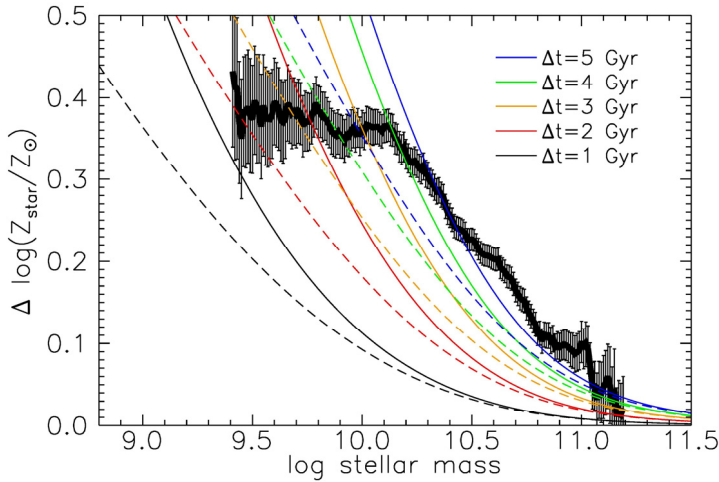

**Extended Data Fig.5**. **Effect of a constant star formation efficiency on stellar metallicity evolution.** As for Fig.2b, but for the case of a constant star formation efficiency of $\varepsilon$=0.5 Gyr$^{-1}$ (i.e. a constant gas depletion timescale of $\tau_{dep}$= 2 Gyr) after strangulation. At $M_{star}$ > ~$10^{10}$ $M_{\odot}$, the observed mass-dependent metallicity enhancement is still consistent of $\Delta t$ ~ 4 or 5 Gyr, while at lower stellar masses it requires a shorter $\Delta t$, as explained in the text. Error bars on the black line indicate the 1σ uncertainty in the metallicity difference.

**References for the Methods:**

31. Peng, Y., Lilly, S. J., Renzini, A. & Carollo, C. M. Mass and Environment as Drivers of Galaxy Evolution. II. The Quenching of Satellite Galaxies as the Origin of Environmental Effects. *Astrophys. J*. **757**, 4 (2012)
32. Blanton, M. R. & Roweis, S. K-Corrections and Filter Transformations in the Ultraviolet, Optical, and Near-Infrared. *Astron. J.*, **133**, 734-754 (2007)
33. Bruzual, G. & Charlot, S. Stellar population synthesis at the resolution of 2003. *Mon. Not. Royal Astron. Soc*. **344**, 1000-1028 (2003)
34. Brinchmann, J., Charlot, S., White, S. D. M., Tremonti, C., Kauffmann, G., Heckman, T. & Brinkmann, J. The physical properties of star-forming galaxies in the low-redshift Universe. *Mon. Not. Royal Astron. Soc*. **351**, 1151-1179 (2004)
35. Gallazzi, A., Charlot, S., Brinchmann, J., White, S. D. M. & Tremonti, C. A. The ages and metallicities of galaxies in the local universe. *Mon. Not. Royal Astron. Soc*. **362**, 41-58 (2005)
36. Lilly, S., Carollo, C. M., Pipino, A., Renzini, A. & Peng, Y. Gas Regulation of Galaxies: The Evolution of the Cosmic Specific Star Formation Rate, the Metallicity-Mass-Star-formation Rate Relation, and the Stellar Content of Halos. *Astrophys. J*. **772**, 119 (2013)
37. Peng, Y. & Maiolino, R. From haloes to Galaxies - I. The dynamics of the gas regulator model and the implied cosmic sSFR history. *Mon. Not. Royal Astron. Soc*. **443**, 3643-3664 (2014)
38. Bigiel, F., Leroy, A., Walter, F., Brinks, E., de Blok, W. J. G., Madore, B. & Thornley, M. D. The Star Formation Law in Nearby Galaxies on Sub-Kpc Scales. *Astrophys. J*. **136**, 2846-2871 (2008)
39. Savaglio, S. et al. The Gemini Deep Deep Survey. VII. The Redshift Evolution of the Mass-Metallicity Relation. *Astrophys. J*. **635**, 260-279 (2005)
40. Maiolino, R. et al. AMAZE. I. The evolution of the mass-metallicity relation at $z$ > 3. *Astron. Astrophys*. **488**, 463-479 (2008)